%% file: main.tex
\documentclass[sigconf]{acmart}
\usepackage{lipsum}
\usepackage{balance}
\usepackage{pifont}
\usepackage{subcaption}
\usepackage[linesnumbered,ruled,slide]{algorithm2e} 

\usepackage{tcolorbox}
\usepackage{lipsum}
\usepackage{multirow}
\usepackage{colortbl} 
\definecolor{ocean}{RGB}{224, 255, 255}
\definecolor{gray}{RGB}{125, 125, 125}

\usepackage{listings}
\usepackage{color}

\definecolor{dkgreen}{rgb}{0,0.6,0}
\definecolor{gray}{rgb}{0.5,0.5,0.5}
\definecolor{mauve}{rgb}{0.58,0,0.82}

\SetCommentSty{mycommfont}

\newcommand{\highlight}[1]{\textcolor{orange}{#1}}
\newcommand{\cached}[1]{\scriptsize{\highlight{#1}}}

\AtBeginDocument{%
  }

\setcopyright{acmlicensed}
\copyrightyear{2018}
\acmYear{2018}
\acmDOI{XXXXXXX.XXXXXXX}

\acmConference[Conference acronym 'XX]{Make sure to enter the correct
  conference title from your rights confirmation emai}{June 03--05,
  2018}{Woodstock, NY}
\acmISBN{978-1-4503-XXXX-X/18/06}

\begin{document}

\newcommand{\name}{{TiInsight}\xspace}
\newcommand{\tisql}{{TiSQL}\xspace}
\newcommand{\tichart}{{TiChart}\xspace}
\newcommand{\papername}{{Towards Automated Cross-domain Exploratory Data Analysis through Large Language Models}\xspace}

\title{\papername}

\settopmatter{authorsperrow=5} 
 
\author{Jun-Peng Zhu}
\affiliation{
  \institution{East China Normal University \& PingCAP, China}
  \country{}
}

\author{Boyan Niu}
\affiliation{
  \institution{PingCAP, China}
  \country{}
}

\author{Peng Cai}
\affiliation{
  \institution{East China Normal University, China}
  \country{}
}

\author{Zheming Ni}
\affiliation{
  \institution{PingCAP, China}
  \country{}
}

\author{Jianwei Wan}
\affiliation{
  \institution{PingCAP, China}
  \country{}
}

\author{Kai Xu}
\affiliation{
  \institution{PingCAP, China}
  \country{}
}

\author{Jiajun Huang}
\affiliation{
  \institution{PingCAP, China}
  \country{}
}

\author{Shengbo Ma}
\affiliation{
  \institution{PingCAP, China}
  \country{}
}

\author{Bing Wang}
\affiliation{
  \institution{PingCAP, China}
  \country{}
}

\author{Xuan Zhou}
\affiliation{
  \institution{East China Normal University, China}
  \country{}
}

\author{Guanglei Bao}
\affiliation{
  \institution{PingCAP, China}
  \country{}
}

\author{Donghui Zhang}
\affiliation{
  \institution{PingCAP, China}
  \country{}
}

\author{Liu Tang}
\affiliation{
  \institution{PingCAP, China}
  \country{}
}

\author{Qi Liu}
\affiliation{
  \institution{PingCAP, China}
  \country{}
}

\renewcommand{\shortauthors}{Jun-Peng Zhu et al.}

\input{src/abstract}



\received{20 February 2007}
\received[revised]{12 March 2009}
\received[accepted]{5 June 2009}

\maketitle

\input{src/introduction}
\input{src/architecture}
\input{src/hdc}
\input{src/tisql}
\input{src/tichart}
\input{src/experiments}
\input{src/related}
\input{src/conclusion}

\balance
\bibliographystyle{ACM-Reference-Format}
\bibliography{ref}

\end{document}

%% file: src/abstract.tex
\begin{abstract}
Exploratory data analysis (EDA), coupled with SQL, is essential for data analysts involved in data exploration and analysis.
However, data analysts often encounter two primary challenges: (1) the need to craft SQL queries skillfully, and (2) the requirement to generate suitable visualization types that enhance the interpretation of query results.
Due to its significance, substantial research efforts have been made to explore different approaches to address these challenges, including leveraging large language models (LLMs).
However, existing methods fail to meet real-world data exploration requirements primarily due to (1) complex database schema; (2) unclear user intent; (3) limited cross-domain generalization capability; and (4) insufficient end-to-end text-to-visualization capability.

This paper presents \name, an automated SQL-based cross-domain exploratory data analysis system. 
First, we propose hierarchical data context (i.e., HDC), which leverages LLMs to summarize the contexts related to the database schema, which is crucial for open-world EDA systems to generalize across data domains.
Second, the EDA system is divided into four components (i.e., stages): HDC generation, question clarification and decomposition, text-to-SQL generation (i.e., \tisql), and data visualization (i.e., \tichart).
Finally, we implemented an end-to-end EDA system with a user-friendly GUI interface in the production environment at PingCAP.
We have also open-sourced all APIs of \name to facilitate research within the EDA community.
Through extensive evaluations by a real-world user study, we demonstrate that \name offers remarkable performance compared to human experts.
Specifically, \tisql achieves an execution accuracy of 86.3\% on the Spider dataset using \mbox{GPT-4}.
It also demonstrates state-of-the-art performance on the Bird dataset.
During six months of public testing, \tisql demonstrates an execution success rate of 82.3\% on real-world EDA tasks.
\end{abstract}

%% file: src/introduction.tex
\section{Introduction}

\subsection{Problem Statement}

Exploratory data analysis (EDA) \cite{chat2query, ma2023insightpilot, ma2023xinsight, milo2020automating, he2024text2analysis}, coupled with SQL, assumes a crucial role for data analysts engaged in data exploration and analysis.
The EDA harnesses SQL to construct queries capable of extracting vital information from databases.
Generally, it involves ``hands-on'' interaction with a dataset, where users iteratively apply analysis actions (e.g., filtering, aggregations, sorting, visualizations), generating statistical insights, and constructing comprehensive views and reports.
The EDA process facilitates a deeper understanding of the data, revealing latent patterns and trends that offer valuable insights to guide subsequent decision-making endeavors.
In particular, the user explores a dataset D = \{databases, schema, tuples, columns\}, where \textit{databases} refer to the data sources the user wants to explore, which may encompass multiple databases; \textit{schemas} refer to the database schema, encompassing tables, indexes, views, and other related elements; \textit{tuples} represent individual records of data, and \textit{columns} represent the attributes of those tuples.
The data analyst then performs a series of data analysis operations $s_1, d_1, s_2, d_2, \ldots, s_n, d_n$, where $s_i$ indicates an SQL statement and $d_j$ represents the visualization of the results. 
The user determines the next steps after executing $s_t$ and $d_t$.
The EDA faced two key technical challenges: (1) proficiency in SQL and (2) generating appropriate visualization types. 

\vspace{-0.3cm}
\subsection{Limitations of Prior Art}
There are many state-of-the-art (SOTA) approaches to address both challenges. 
Specifically, regarding SQL proficiency, there are numerous SOTA text-to-SQL approaches, which effectively reduce the complexity of crafting SQL queries for data analysts.
However, in the EDA context, SOTA text-to-SQL approaches still have their limitations:

\noindent\textbf{(1) Complex database schema}. In real-world EDA scenarios, the data to be explored is typically stored in databases with complex schema \cite{avrilia2024nl2sql, li2024codes, zhang2024finsql}. For example, in financial data analysis, each database contains numerous wide tables with hundreds of columns \cite{zhang2024finsql, avrilia2024nl2sql}. 
The schema complexity far exceeds that of existing benchmarks, such as Spider \cite{spider}.	
Existing text-to-SQL methods struggle to handle such complex scenarios, as evidenced by low accuracy \cite{zhang2024finsql} on this dataset.
Moreover, large language model-based methods must consider context window limits \cite{gpt4, chatgpt2022, touvron2023llama} when constructing prompts, significantly impacting accuracy in complex scenarios.
In the worst-case scenario, a query involving numerous tables and hundreds of columns may exceed the context window limits during prompt construction, preventing the method from functioning properly.
On the other hand, the large number of tables and columns increases the context length, which can significantly decrease accuracy.
In real-world customer scenarios at PingCAP, users define numerous abbreviations that may refer to a business, a location, or other entities.	
This further increases the complexity of the schema, requiring more domain knowledge.

\noindent\textbf{(2) Unclear user intent}. In real-world EDA scenarios, users often find it challenging to express their thoughts in natural language, which makes it difficult to convey their intent \cite{he2024text2analysis, avrilia2024nl2sql}.
This challenge is evident in users' inability to formulate clear intent parameters.
For example, in the customer scenarios of PingCAP, customers often ask, ``What is the growth rate?''.
This query may lack a time parameter; a more precise formulation would be, ``What is the growth rate for the current year?''.
Additionally, user queries frequently include abbreviations \cite{zhang2024finsql}; for example, ``DoD analysis for daily bills.''.
In the worst-case scenario, users may struggle to clearly articulate their intent at the beginning of a data analysis task.
In many cases, the user's natural language (NL) intentions cannot be captured by a single SQL statement \cite{pourreza2024din, wang2024macsql, avrilia2024nl2sql}.
Accurately addressing these queries requires not only the semantic parsing capabilities of large language models but also robust data analysis skills to infer intent beyond the explicit query \cite{he2024text2analysis}.
Existing SOTA text-to-SQL approaches struggle to generate SQL in this context, let alone address accuracy issues \cite{gao2023text, pourreza2024din, avrilia2024nl2sql}.

\noindent\textbf{(3) Limited cross-domain generalization}. Existing text-to-SQL methods \cite{zhang2024finsql, li2024codes} are designed for specific domains, which results in poor execution accuracy when applied to a different data domain, requiring fine-tuning for each domain.
In an end-to-end enterprise-grade system, this becomes virtually impossible.	
Furthermore, fine-tuning introduces substantial overhead in terms of time, space, and cost.
Consequently, integrating SOTA text-to-SQL approaches into an end-to-end EDA system proves challenging.
For example, in the customer scenarios of PingCAP, there are applications in finance, retail, and other industries, but SOTA methods lack cross-domain generalization capabilities.
Fine-tuning for different business contexts is required, making it impractical.	
On the other hand, fine-tuning requires well-labeled data \cite{zhang2024finsql}, but many industries lack adequate labeled datasets.
This further restricts the adaptability of fine-tuning methods in text-to-SQL.
\textit{It is important to note that current SOTA methods are heavily benchmark-oriented and perform poorly in real-world cross-domain text-to-SQL challenges.}

Beyond these limitations in text-to-SQL, there are also challenges in data visualization for EDA systems:

\noindent\textbf{(1) Insufficient end-to-end automated text-to-visualization capability}.
The existing EDA systems typically offer three primary visualization methods: \ding{172} manual visualization \cite{d3js, tableau, powerbi, stolte2002polaris}, where the user selects visualization types for a given dataset on selected attributes. \ding{173} natural language description \cite{tableau,he2024text2analysis}, where the user specifies visualizations, such as ``help me create a line chart to visualize the sales of PC machines''. These methods typically either specify the type of visualization in text or directly generate visualizations from text, whereas our scenario requires a process involving text-to-SQL execution followed by data-to-chart generation. \ding{174} table-to-chart \cite{zhou2021table2charts, luo2018deepeye}, where the system automatically generates appropriate charts based on the dataset, a process that is often highly complex.
The majority of research in this area concentrates on exploring multi-dimensional data.
Most importantly, these methods generally visualize attributes pre-selected by the user or autonomously suggest interesting attributes from the dataset.
This approach differs significantly from our scenario.

\noindent\textbf{(2) Overly complex table-to-chart recommendation process}.
Existing methods \cite{zhou2021table2charts, luo2018deepeye, vartak2015seedb} frequently employ complex algorithms, such as reinforcement learning \cite{zhou2021table2charts}, significantly increasing system complexity.
Some research \cite{wongsuphasawat2017voyager, lee2021lux} requires users to master specific visualization query languages, which significantly restricts the applicability of these methods in end-to-end systems.
In numerous user studies, we observed that simple visualization types, such as pie, bar, and line charts, were preferred during data exploration, while more complex visualizations hindered users' ability to extract insights.
In particular, data analysts have accumulated numerous heuristics rules for data visualizations, but these lack a straightforward way of applying them to visualization recommendations.
The advent of LLMs presents new opportunities for rule-based visualization.

\vspace{-0.3cm}
\subsection{Key Technical Challenges}
Enabling an end-to-end SQL-based automated EDA system offers significant benefits for data exploration and analysis.
However, there is no free lunch in this context.	
There are three major \textbf{challenges} that need to be addressed:

\noindent\textbf{C1. How to mitigate the impact of complex database schema and unclear user intent?} 
This is primarily due to the prevalence of complex database schema and ambiguous user intentions in real-world EDA scenarios, significantly reducing EDA task accuracy.
However, addressing these challenges without incurring additional costs remains highly difficult.

\noindent\textbf{C2. How to improve generalization and adaptability across different data domains?} 
In industrial practice at PingCAP, various business data scenarios, such as finance and retail, are commonly encountered.
EDA methods are typically tailored to specific data domains, leading to the absence of a universal EDA system that can adapt to diverse scenarios.
Therefore, developing automated EDA systems that provide generalization and adaptability remains a significant challenge.

\noindent\textbf{C3. How to design an automated EDA system that translates NL into visualization results with minimal human intervention?}
Efficient end-to-end exploratory analysis requires the design of a dedicated system.
The complexity of designing EDA systems is amplified by uncontrollable natural language inputs, unknown exploratory datasets, and complex domain-specific knowledge.
It is a key research focus to investigate a feasible automated EDA pipeline in data analysis.

\subsection{Proposed Approach}

In this paper, we propose \name, a SQL-based enterprise-grade multi-stage automated cross-domain exploratory analysis system through large language models.
To address the first two challenges, we propose HDC, a hierarchical data context approach that leverages LLMs to summarize the database schema context, which is vital for open-world EDA systems to generalize across exploratory data domains.
First, \name summarizes database-related context information using HDC, such as column descriptions and table summaries. The primary data entities representing the database are subsequently abstracted. Ultimately, a database summary is generated that encompasses the entity descriptions, the tables associated with those entities, the key attributes of the entities, and the columns to which these entities may be mapped. 
These data contexts are stored in vector databases (e.g., ChromaDB \cite{chroma}, Pinecone \cite{pinecone}, TiDB Vector \cite{tidb_vector}).
Drawing on this data context, \name also recommends several exploratory questions to facilitate the user's understanding of the data.

\name is divided into four components (i.e., stages): HDC generation, question clarification and decomposition, text-to-SQL generation (i.e., \tisql), and data visualization (i.e., \tichart).
Subsequently, \name processes the natural language questions posed by users.	
Initially, it clarifies the user questions within the hierarchical data context, generating a precise data exploration question. 
It then determines whether the question can be answered with a single SQL statement.	
If not, it decomposes the complex question into multiple sub-questions and provides a detailed description of each sub-question.
These clarification questions are subsequently provided to the \tisql, which generates the corresponding SQL statement and integrates it with the self-refinement chain for further improving text-to-SQL accuracy.
Finally, \name generates the visualization chart utilizing the integrated \tichart.
An end-to-end EDA system with a user-friendly GUI interface has been developed and deployed in the production environment at PingCAP. 

\subsection{Key Contributions and Outline}
To summarize, this paper makes the following contributions:

(1) We investigate how to design key components in an SQL-based automated EDA system and analyze why existing methods are inadequate for enterprise-grade automated EDA systems.

(2) We propose a hierarchical data context approach called HDC.
\name with HDC is designed to facilitate data exploration of complex enterprise-grade datasets and enable cross-domain data analysis.
Furthermore, we introduce a map-reduce framework into \name to facilitate large-scale and complex database schemas during the HDC generation.
The generated data context is stored in vector databases.

(3) Building on HDC, we propose a cross-domain text-to-SQL method named \tisql. The \tisql explores various prompting techniques to enhance the accuracy of text-to-SQL generation. Building on this foundation, we introduce \tichart, which implements a rule-based visualization recommendation.

(4) We implement and deploy an automated end-to-end EDA system with a user-friendly GUI interface in the production environment at PingCAP.
Simultaneously, we have also open-sourced all APIs of \name to facilitate research within the EDA community.

(5) We evaluate the performance of \name. The experimental results demonstrate that \name achieved remarkable performance.
In particular, \tisql achieves 86.3\% execution accuracy in the Spider dataset using GPT-4 without requiring additional model training or fine-tuning.
It also demonstrates state-of-the-art performance on the Bird dataset.
During six months of public testing, \tisql has achieved an execution success rate of 82.3\% on real-world EDA tasks.

The system is available at \textcolor{blue}{\url{https://tiinsight.chat}}.
The demo video is available at \textcolor{blue}{\url{https://youtu.be/01eYVmuu6uI}}.
The experimental evaluation code of \tisql is available at \textcolor{blue}{\url{https://github.com/tidbcloud/tiinsight}}.
The RESTful interface definition of APIs for \name is available at \textcolor{blue} {\url{https://docs.pingcap.com/tidbcloud/use-chat2query-api}}.

The remainder of the paper is organized as follows.
The architecture of \name is discussed in Section~\textcolor{blue}{\S \ref{sec:overview_arch}}.
We give the details of the hierarchical data context and question clarification and decomposition in Section~\textcolor{blue}{\S \ref{sec:hdc}}.
The detailed implementation of \tisql and \tichart is presented in Section~\textcolor{blue}{\S \ref{sec:tisql}} and Section~\textcolor{blue}{\S \ref{sec:tichart}}, respectively. 
The experimental evaluation is presented in Section~\textcolor{blue}{\S \ref{sec:experiment}}.
We review related work in Section~\textcolor{blue}{\S \ref{sec:relate}} and conclude in Section~\textcolor{blue}{\S \ref{sec:conclusion}}.

%% file: src/architecture.tex
\section{The Overview of \name} \label{sec:overview_arch}

\begin{figure}[!t]
\centering
\includegraphics[width=0.40\textwidth]{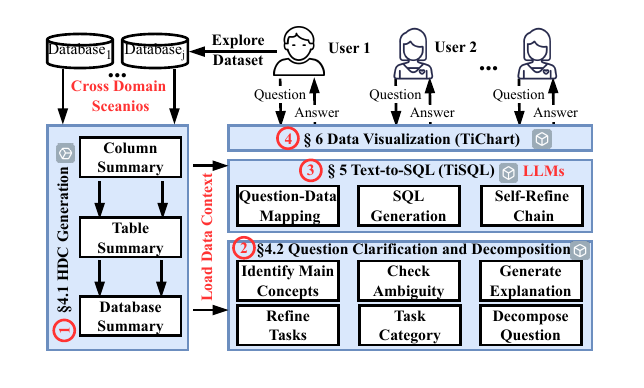}
\vspace{-0.3cm}
\caption{Overall architecture of \name.}
\vspace{-0.3cm}
\label{fig:architecture}
\end{figure}

This section gives a high-level introduction of \name. As shown in Figure~\ref{fig:architecture}, \name has four functional components to facilitate cross-domain exploratory data analysis tasks.

\noindent\textbf{\underline{\ding{172} HDC generation}}. We perform cross-domain exploratory data analysis. To address \textbf{C1} and \textbf{C2}, we design HDC generation, which uses LLMs to generate the hierarchical data context of the database schema efficiently. The LLMs are powerful for summarizing text content from massive datasets.
It can capture the most significant content from the database schema to guide the subsequent exploratory process.
This component allows us to interpret the data provided by the user across various hierarchies of the database schema.
In Section~\textcolor{blue}{\S \ref{subsec:hdc}}, we introduce how to summarize the information provided by the database schema from different perspectives, including columns, tables, and databases, and how to address complex database schemas of real-world scenarios.

\noindent\textbf{\underline{\ding{173} Question Clarification and Decomposition}}. We decode and clarify the questions of different users. To address \textbf{C1} and \textbf{C2}, we further implement a solution to clarify questions using large language models. Specifically, this component resolves ambiguous user intentions and provides possible explanations. Subsequently, we refine the current user task. Simultaneously, \name assesses whether the user task should be decomposed into a series of sub-tasks. Section~\textcolor{blue}{\S \ref{subsec:qdc}} introduces how large language models can effectively clarify and decompose tasks.

\noindent\textbf{\underline{\ding{174} Text-to-SQL (\tisql)}}. We employ \tisql to translate user questions into SQL statements. First, \tisql needs to map the user question to specific tables and columns while providing the values for SQL conditional statements.
In this process, we propose both coarse-grained and fine-grained mapping methods.
Subsequently, \tisql employs the large language model to generate SQL statements.
We carefully design the prompt of \tisql. However, the SQL statements generated in this step still contain errors.	Finally, \tisql employs a self-refinement chain to enhance the generated SQL.
In Section~\textcolor{blue}{\S \ref{sec:tisql}}, we elaborate on using a large language model combined with prompt engineering to generate SQL statements without the need for fine-tuning efficiently.

\noindent\textbf{\underline{\ding{175} Data Visualization (\tichart)}}. In \name, we utilize \tichart to visualize the user's query results to the fullest extent. A key challenge in the data visualization process arises from the complexity of user tasks, which can generate numerous sub-tasks, leading to uncertainty in the final data query (e.g., attributes and data).
This greatly amplifies the complexity of visualizing the resulting data.
To address this challenge, \tichart combines a rule-based approach with a large language model to generate clear and easy-to-understand visualizations.
In Section~\textcolor{blue}{\S \ref{sec:tichart}}, we provide a detailed discussion on how rules can be integrated into large models to improve the generation of visualization results.

The above architecture solves \textbf{C3} by considering the complexity of different exploratory scenarios.
Specifically, we propose the hierarchical data context approach to address the challenges that current EDA scenarios cannot operate across domains.
Furthermore, our proposed HDC generation approach facilitates the implementation of an end-to-end cross-domain text-to-SQL tool (i.e., \tisql) and a data visualization tool (i.e., \tichart). 
We have open-sourced all APIs from various components to facilitate research within the EDA community.

%% file: src/hdc.tex
\vspace{-0.3cm}
\section{Hierarchical Data Context and Question Clarification and Decomposition}  \label{sec:hdc}

In this section, we first present a detailed implementation of the hierarchical data context (\textcolor{blue}{\S \ref{subsec:hdc}}). Then, we give the design for the clarification and decomposition of the questions (\textcolor{blue}{\S \ref{subsec:qdc}}). 

\begin{figure}[!t]
\centering
\includegraphics[width=0.23\textwidth]{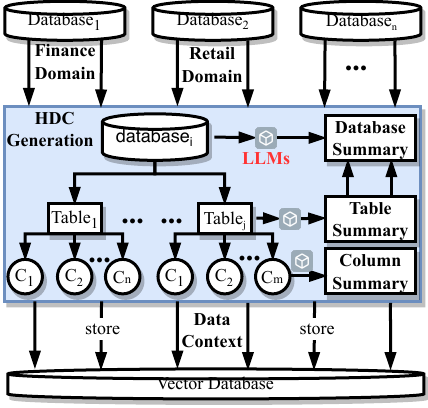}
\vspace{-0.3cm}
\caption{Hierarchical data context generation of \name.}
\vspace{-0.3cm}
\label{fig:hdc}
\end{figure}

\begin{algorithm}[!t]
\SetAlgoVlined
\SetKwFunction{FMain}{Map}
\SetKwProg{Fn}{Function}{:}{}
\Fn{\FMain{table, t}}{
\label{alg:table:map:begin}
    batch\_table\_description = [ ]\;
    blocks = split(table)\;
    \For{b in blocks}{
        batch\_column\_description = retrieve\_column\_description(b)\;
        mPrompt = preparePrompt(batch\_column\_description, table)\; \label{alg:table:prompt}
        \tcc{It can be processed in parallel using multiple threads.}
        mResult = LLM(mPrompt, t)\;
        batch\_table\_description.append(mResult)\;
    }
    \Return batch\_table\_description\;
    \label{alg:table:map:end}
}

\SetAlgoVlined
\SetKwFunction{FMain}{Reduce}
\SetKwProg{Fn}{Function}{:}{}
\Fn{\FMain{batch\_table\_description, table, t, max\_tokens}}{ \label{alg:table:reduce:begin}

    \If{len(batch\_table\_description) == 1}{
        rPrompt = preparePrompt(batch\_table\_description[0], table)\;
        \Return LLM(rPrompt, t)\;
    }
    
    new\_batch\_table\_description = [ ]\;
    current\_batch = [ ]\;
    \For{description in batch\_table\_description}{
         \If{get\_token\_count(current\_batch) + get\_token\_count(description) > max\_tokens}{
            rPrompt = preparePrompt(current\_batch, table)\;
            reduced\_result = LLM(rPrompt, t)\;
            new\_batch\_table\_description.append(reduced\_result)\;
            current\_batch = [ ]\;
         }
        current\_batch.append(description)\; 
    }
    \If{current\_batch}{
        rPrompt = preparePrompt(current\_batch, table)\;
        reduced\_result = LLM(rPrompt, t)\;
        new\_batch\_table\_description.append(reduced\_result)\;
    }
    \Return Reduce(new\_batch\_table\_description, table, t, max\_tokens)\; \label{alg:table:reduce:end}
}

\SetAlgoVlined
\SetKwFunction{FMain}{MRTableDes}
\SetKwProg{Fn}{Function}{:}{}
\Fn{\FMain{table, t, max\_tokens}}{
    
    mResult = Map(table, t)\;
    rResult = Reduce(mResult, table, t, max\_tokens)\;
   
    save\_table\_description(table, rResult)\; \label{alg:table:mr:end}
 }
\caption{Table Description Map-Reduce Framework.}
\label{alg:table}
\LinesNumbered
\end{algorithm}

\subsection{Hierarchical Data Context} \label{subsec:hdc}

The hierarchical data context (i.e., HDC) forms the foundation of \name, with all subsequent component designs built upon it.
The hierarchical data context is designed primarily to improve the understanding of the database schema, which includes column information, table information, table-to-table relationships, and overall database details.
This design is derived from experience during the proof-of-concept (POC) process at PingCAP.
In PingCAP's data analysis business operations, data analysts first collect and organize relevant database schemas information when encountering unfamiliar datasets to gain a better understanding of the database.
We leveraged this experience and fully automated the process using LLMs.
As illustrated in Figure~\ref{fig:hdc}, the HDC comprises three components: column summary (\textcolor{blue}{\S \ref{subsec:col}}), table summary (\textcolor{blue}{\S \ref{subsec:table}}), and database summary (\textcolor{blue}{\S \ref{subsec:database}}). 
Due to space constraints, the prompt used in this section is not included in the paper.

\subsubsection{Column Summary}  \label{subsec:col}
Columns (a.k.a., fields or attributes) are the fundamental components of a database schema.
In real-world scenarios, two main challenges arise with columns: (1) column names are often abbreviated, and (2) there can be a very large number of columns, sometimes reaching thousands, and each column contains a lot of data. 
To address the first technical challenge, we propose building an embedding service to store domain knowledge (a.k.a., terms) in a vector database. When the LLMs generate a summary of column descriptions, they first retrieve relevant domain knowledge from the vector database, dynamically constructing prompts as supplementary explanations.

A column-by-column summarization approach would result in high latency due to repeated calls to the large language model.
Additionally, using the LLM for each column individually is inefficient, leading to significant token wastage and increased costs.
To address the second challenge,  we propose a parallel, grouping-based approach to mitigate this issue by vertically partitioning the table. 
First, HDC groups columns into fixed sets (i.e., vertical partitioning); empirically, groups contain $40$ columns for GPT-4 and $80$ columns for other models.
This limitation is primarily due to the context window constraints of these models.
Each group is subsequently allocated threads from a thread pool before the LLM initiates the column description process. 
Comments may be included in the schema information provided by the schema for each column.
For example, the \textit{date} column might include a description of the date format, such as YYYY-mm-dd.
After the aforementioned process, we append the column comments to each column description.
Additionally, for each column, we randomly select several rows (i.e., three rows in this paper) to include in the column description, aiding the LLM in understanding the column's value type and other relevant information. 
Finally, the LLM receives the column content and dynamically constructs the CoT prompt to facilitate the column summary.

Our approach utilizes significantly fewer tokens to concisely convey column-related information compared to methods that encode all tables in the prompt \cite{zhang2023reactable}.
This distinction is particularly critical for tables that contain millions of rows of data.

\subsubsection{Table Summary} \label{subsec:table}

Next, we utilize the column summary obtained in Section~\textcolor{blue}{\S\ref{subsec:col}} to summarize the information for each table in a table-by-table manner.
The table summary consists primarily of two components: the table description and the table relationship.

\begin{figure}[!t]
\centering
\includegraphics[width=0.30\textwidth]{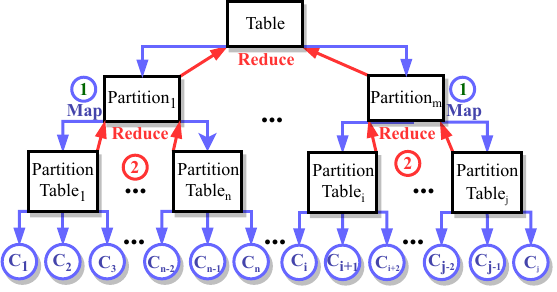}
\vspace{-0.3cm}
\caption{An illustration of map-reduce for table description.}
\vspace{-0.3cm}
\label{fig:mapreduce}
\end{figure}

\noindent\textbf{\underline{Table Description}}.
In real-world scenarios, tables may contain thousands of columns, each potentially with millions of rows.
Encoding the table data within a prompt is impractical, so we encode only the column summaries.
However, large and wide tables cannot encode all columns in a single prompt, as this often exceeds the context window limits of LLMs.
Consequently, context window limits are a significant challenge in table summarization for LLMs.
In this paper, we propose a map-reduce framework to summarize large tables. 
Algorithm~\ref{alg:table} provides a detailed implementation of the table summary obtained using the map-reduce framework.
First, it divides the wide table into multiple blocks using a vertical partitioning of columns.
Then, it creates a summary for each block (lines~\ref{alg:table:map:begin}-\ref{alg:table:map:end}).
In this process, we retrieve the column descriptions corresponding to each block from the vector database.
The prompt is constructed dynamically using this information (line~\ref{alg:table:prompt}).
Finally, it uses the reduce mode to generate the table description (lines~\ref{alg:table:reduce:begin}-\ref{alg:table:reduce:end}).
Additionally, we save the context information of the table description obtained from the exploration to the vector database (line~\ref{alg:table:mr:end}).
Figure~\ref{fig:mapreduce} illustrates the above process.
The map-reduce framework offers two key benefits: (1) It prevents exceeding the context window limits of the LLM, and (2) It supports multi-threads in parallel summarization across different stages.

To create a comprehensive table description, we need to gather additional information, including the primary key, key attributes, table type, table entity, and natural language description.
Specifically, the LLM may identify multiple primary keys, and we prompt it to select the most likely one.
This primary key should better reflect the importance and uniqueness of the table.
Key attributes play a vital role in understanding the purpose and content of the table.
We prompt the LLM to identify and list a maximum of five key attributes from the table presented in this paper.
In addition, we categorize the table type as either dimension, bridge, or fact.
This classification is especially crucial for the OLAP analysis.
Finally, we identify the main entity that the table focuses on.

\begin{algorithm}[!t]
\SetAlgoVlined
\SetKwFunction{FMain}{TSTabRel}
\SetKwProg{Fn}{Function}{:}{}
\Fn{\FMain{table, t, similar\_count}}{
    \tcc{Stage 1: Coarse-grained similarity search from vector database.}
    doc = retrieve\_table\_description(table)\;
    relevant\_tables, relevant\_tables\_description = coarse\_grained\_search(doc, similar\_count)\;

    \tcc{Stage 2: Fine-grained explore the results from stage 1 using LLM.}
    prompt = (table, table.columns, relevant\_tables\_description)\;
    rel = LLM(prompt, t)\;
    \tcc{Save the table relationship to the vector database.}
    save\_table\_relationship(table, rel);
 }
\caption{Table Relationship.}
\label{alg:tl}
\LinesNumbered
\end{algorithm}

\noindent\textbf{\underline{Table Relationship}}. Table relationships identify all referential integrity relationships between a specified table and others, which helps to optimize complex queries. 
We need to identify the following elements: (1) the referencing table, (2) the referenced table, (3) the foreign key column, (4) the primary key column, and (5) the nature (i.e., relationship type) of the inferred relationship corresponding to the foreign key column (e.g., 1: 1, 1: N).
Accurately identifying these relationships in real-world scenarios remains challenging due to the large number of tables and the complexity of their relationships. 
For example, in PingCAP’s business, a single database may have thousands of tables, each containing thousands of columns.

In this paper, based on the table descriptions, we propose a two-stage approach for identifying the table relationships to address the above challenges, as shown in Algorithm~\ref{alg:tl}. 
To determine the referential integrity of a specific table with other tables, in \textbf{stage~1 (i.e., coarse-grained search)}, we retrieve \textit{similar\_count} tables from the vector database using the specified table description (lines 2-3). Typically, the number of \textit{similar\_count} is set to $20$. 
In \textbf{stage~2 (i.e., fine-grained explore)}, we input the specified table along with the table retrieved in stage~1 into the LLMs for fine-grained identification by CoT prompt (lines 4-5).
We iterate over each table to establish an integrity relationship for all tables within the database. Given $n$ tables, the time complexity to explore all tables for each specified table is $O(n^2)$. Using a coarse-grained search reduces this complexity, allowing us to avoid exploring every table and achieve a time complexity of $O(n)$.

\subsubsection{Database Summary} \label{subsec:database}

\begin{algorithm}[!t]
\SetAlgoVlined
\SetKwFunction{FMain}{Entity}
\SetKwProg{Fn}{Function}{:}{}
\Fn{\FMain{schema, t, N}}{
   \tcc{Step 1: Select the top N tables with the highest number of relationships.}
   candidate\_tables = topN(schema, N)\;
   \tcc{Step 2: Extract the entities from candidate\_tables using LLM.}
   prompt = preparePrompt(candidate\_tables, candidate\_tables\_summary)\;
   entities = LLM(prompt, t)\;
   \tcc{Step 3: Save the entities to the vector database.}
   save\_entity\_context(database, entities)\;
 }
\caption{Entity Extraction.}
\label{alg:ee}
\LinesNumbered
\end{algorithm}

To facilitate efficient data exploration across multiple databases, we propose extracting representative \textbf{entities} for each database, enabling a \textbf{database summary} based on these abstracted entities. 
A database consists of numerous tables with complex relationships. 
Following the influence maximization principle \cite{li2018influence}, a table with a higher number of relationships to other tables is considered more important within the database.
Conversely, for a table with very few relationships, its properties are generally insignificant.
Therefore, we propose an entity extraction method that leverages the number of relationships between tables, as outlined in Algorithm~\ref{alg:ee}. 
The algorithm first selects the top N tables (line 2) with the highest number of relationships from the current database. 
In the PingCAP business, the value of $N$ is typically set to $20$.
A prompt is then constructed using these tables and their summaries, and the LLM is employed to infer their entities (lines 3-4). 
Finally, the inferred entities are saved in the vector database (line 5).

For the extracted entity, we employ a CoT prompt with the LLM to infer: (1) a larger entity name that accurately represents the grouped tables and reflects their theme; (2) a larger entity summary emphasizing its purpose and the insights it contributes to a holistic understanding of the data source; (3) key attributes related to the larger entity, ensuring they are derived from the key attributes of the grouped tables; and (4) a list of the names of tables involved in the larger entity.

Finally, we infer the database summary from the entities using the LLM with a CoT prompt, which includes the following information:

\begin{itemize}
    \item \textbf{Purpose}: Infer the purpose of the data source. Do not include details about entities.
    \item \textbf{Domain}: Determine the domain of the data source.
    \item \textbf{Business Impact}: Describe in simple terms how the data source aids in business operations, analytics, or decision-making processes.
    \item \textbf{Real-World Example}: Provide an example of real-world scenarios in which the data source can play a pivotal role.
    \item \textbf{User-Friendly Description}: Combine the results from the previous steps into a user-friendly description. Do not emphasize what it includes in the description.
    \item \textbf{Summary}: Write a summary of the data source based on the user-friendly description, highlighting key points, and in no more than $10$ words.
\end{itemize}

We save the above information to a vector database and generate various suggestion questions for data analysis using the LLM with a CoT prompt, combining the entity and database descriptions.
These questions correspond to the following types of data analysis: descriptive, inferential, diagnostic, predictive, and prescriptive.
These questions assist users in exploring the dataset from the ground up.

\vspace{-0.3cm}
\subsection{Question Clarification and Decomposition}  \label{subsec:qdc}

\noindent\textbf{\underline{Question Clarification}}. Upon data import into \name, our system initiates a rapid analysis to establish a hierarchical data context, enabling the user to explore the dataset effectively.
A key challenge for \name is that user questions can be semantically unclear.
For example, in the user scenario of the PingCAP, a common question like ``Which product is the best?'' may refer to sales, customer satisfaction, or other metrics.
To address this ambiguity, we propose using a hierarchical data context to clarify user questions and establish clear exploration intentions.
Another key issue is that user questions may contain abbreviations, which can impact \name to explore the dataset accurately.
To address this issue, \name establishes an embedding service that stores domain-specific knowledge within a vector database.
The component combines this information to dynamically construct prompts and complete the clarification of the question using a CoT prompt.

The main steps for question clarification are: \textbf{(1) Identifying main concepts}: analyze the task context to determine the primary concepts or variables mentioned. \textbf{(2) Checking for ambiguity}: review the identified concepts or variables to assess potential ambiguity based on the hierarchical data context in Section~\textcolor{blue}{\S \ref{subsec:hdc}}. \textbf{(3) Generating explanations}: for each unclear concept or variable, generate a relevant explanation grounded in the provided information. \textbf{(4) Refining the task}: use the generated explanations to clarify the task, ensuring it communicates the intended outcome and resolves any ambiguities. \textbf{(5) Providing a detailed description}: provide a detailed description of the refined task, including the objective, expected outcomes, and any relevant considerations or requirements.

\noindent{\textbf{\underline{Question Decomposition}}}. 
Real-world questions are often too complex for \name to answer fully in a single step, requiring the question to be broken down into sub-questions.
A key challenge in decomposition for LLM is the lack of domain-specific knowledge.
To address this, we propose an approach based on standard operating procedure (SOP) and few-shot examples by similarity-based retrieval to guide LLMs in generating more effective decomposition plans.	
In the PingCAP business scenario, SOP represents a unified logic derived from specific business operations, enabling LLMs combined with SOP to generate more effective sub-tasks.
Building on this, we select decomposition task results with high user satisfaction as few-shot examples to further improve the performance of \name in decomposition tasks.

%% file: src/tisql.tex
\vspace{-0.3cm}
\section{\tisql: Text-to-SQL with Self-refine Chain} \label{sec:tisql}

\begin{algorithm}[!t]
\SetAlgoVlined
\SetKwFunction{FMain}{TabColFilter}
\SetKwProg{Fn}{Function}{:}{}
\Fn{\FMain{clarified\_task, N}}{
    relevant\_tables = [ ]\;
    \tcc{The variable includes tables and columns.}
    necessary\_tables\_columns = [ ]\;
   \tcc{The coarse-grained retrieves the N-related tables.}
   relevant\_tables = retrieves(clarified\_task, N)\;
    \tcc{Split relevant\_tables into multiple blocks.}
   blocks = split(relevant\_tables)\;
   map\_result = Map(blocks, clarified\_task)\;

   necessary\_tables\_columns = Reduce(map\_results)\;
   \Return necessary\_tables\_columns\;
 }

\SetKwFunction{FMain}{Map}
\SetKwProg{Fn}{Function}{:}{}
\Fn{\FMain{blocks, clarified\_task}}{
    results = [ ]\;
    \For{block in blocks}{
        \tcc{Each block is filtered through the LLM  in parallel in separate threads.}
        prompt = preparePrompt(block.HDC, clarified\_task)\;
        result = LLM(prompt, t)\;
        results.append(result)\;
   }
   \Return results\;
}

\SetKwFunction{FMain}{Reduce}
\SetKwProg{Fn}{Function}{:}{}
\Fn{\FMain{map\_result}}{
    necessary\_tables\_columns = [ ]\;
    \tcc{Reduce the result of each block.}
    \For{result in map\_results}{
        \If{result not in necessary\_tables\_columns}{
            necessary\_tables\_columns.append(result);
        }
    }
    \Return necessary\_tables\_columns\;
}
\caption{Tables and Columns Filter.}
\label{alg:sf}
\LinesNumbered
\end{algorithm}

In real-world scenarios, a database typically includes thousands of tables, each with hundreds or thousands of columns, and each table holds millions of records.
Encoding all of this information into a prompt would exceed the context window limits of LLM.
Therefore, in text-to-SQL tasks, it is essential to streamline the database schema, as an excessive number of schemas can cause errors during the schema-linking \cite{li2023resdsql} process.
To address these challenges, we propose a schema filtering framework based on the map-reduce paradigm to filter tables and columns.
The schema filtering algorithm is presented in Algorithm~\ref{alg:sf}.

As shown in Algorithm~\ref{alg:sf}, \tisql begins with a coarse-grained search, retrieving the top $N$ relevant tables from the vector database based on the clarified question (i.e., clarified\_task) and cosine similarity (line 4). In PingCAP practice, this parameter is set to $30$.
Then, \tisql performs fine-grained filtering to exclude irrelevant tables and columns.	
To prevent the associated table and column summaries from exceeding the LLM's context limit, \tisql divides these tables into more granular groups of related tables (line 5).
Subsequently, map-reduce is employed to process the corresponding block utilizing idle threads (lines 9-21).
In the map phase, the prompt (line 12) is dynamically constructed using the table and column summaries of the current group and the clarified task.
The LLM, combined with the CoT prompt, is then used to identify the relevant tables and required columns (line 13). 
In the reduce phase, we apply the reduce mode for each fine-grained table group to merge the retrieved tables and columns (lines 16-21).
\tisql utilizes both the hierarchical data context (i.e., HDC) and the map-reduce framework to streamline the LLM prompt, significantly reducing the complexity of schema linking.
In particular, \tisql does not separate the table and column recall into two stages.
This approach allows \tisql to leverage additional context from table and column summaries to better map the exploratory tasks of users.

After retrieving the relevant tables and columns, \tisql enters the SQL generation phase, where it uses these tables and columns, along with question-SQL few-shot examples from the vector database, to build the CoT prompt and generate SQL statements dynamically.  
The specific steps include: (1) \textbf{Identify keywords}: Extract all relevant keywords associated with tables and columns from the clarified task.
(2) \textbf{Map keywords to columns and tables}: Match the extracted keywords to specific columns and tables within the schemas, determining the appropriate column and table names of each keyword.
(3) \textbf{Generate SQL query}: Using the results of the mapping in step (2) and the relationships in the identified table, construct an SQL query that meets the task requirements. If no SQL query can be generated, leave the output empty.
(4) \textbf{Rewrite SQL query}: Following the generation of the SQL query in step (3), refine it using rewriting rules: apply the \textit{table.column} format to prevent ambiguity when multiple tables are included, verify that table and column references are accurate and correspond to the correct tables, and enclose all table and column names in backticks to ensure proper referencing and prevent syntax errors.

\begin{figure}[!t]
\centering
\includegraphics[width=0.25\textwidth]{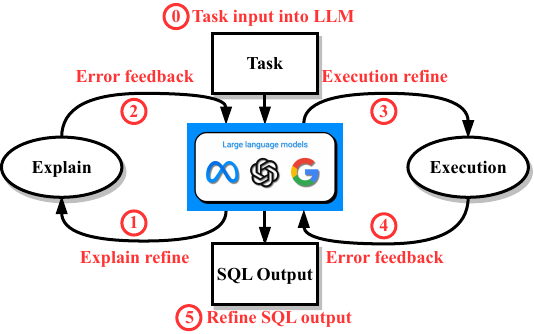}
\vspace{-0.3cm}
\caption{The self-refine chain of \tisql.}
\vspace{-0.3cm}
\label{fig:refine}
\end{figure}

The SQL statements generated from the above procedure may still contain errors; therefore, we propose a self-refinement chain, as illustrated in Figure~\ref{fig:refine}.
After generating SQL statements from the LLM, the results are initially fed into an explain refine process (\ding{172}), where potential errors can be identified (\ding{173}).
The \textit{EXPLAIN} statement is a widely used tool in databases that provides feedback on potential issues in SQL statements without having to execute them.
For example, attempting to parse a non-existent table, view, or column results in an error from the database.	
Similarly, an invalid expression or an incorrect parameter in a query can lead to an error in the \textit{EXPLAIN} output.
For example, executing ``EXPLAIN SELECT * FROM users WHERE id = `abc'::integer;'' returns the error ``invalid input syntax for integer: `abc'.''.
Such errors are frequently encountered in real-world scenarios.
This method enables the identification and enhancement of SQL statements without the need to execute the query.
After refining the SQL statements using the explain statement, certain errors may only become apparent during execution; therefore, we integrate execution refinement into the \tisql refinement chain (\ding{174}).
Execute the SQL statements and relay any feedback from error information to the LLM for correction (\ding{175}).
The final SQL output is generated after iterating the self-refinement chain (\ding{176}).
This process prompts the LLM to self-reflect and learn from its errors, enhancing the quality of its output without requiring additional training on a wide range of text-to-SQL tasks.

%% file: src/tichart.tex
\vspace{-0.3cm}
\section{\tichart: Rule-Based Visualization}  \label{sec:tichart}

Data visualization is a key component of \name, aiding users in comprehending data exploration results more effectively.
In previous data analysis systems, most users needed to use d3js \cite{d3js} and other tools for data visualization.
However, these systems lacked an end-to-end solution with minimal human intervention.	
To address these limitations, we propose a rule-based data visualization approach.
In the PingCAP business scenario, we collected chart visualization preferences of users and combined these rules with LLM recommendations to generate visual output.

First, we identify the data type of each column by analyzing the sample data results.	
The type of data is a crucial factor in determining both visualization methods and accuracy.
Different data types are suited to various visualization methods, each capable of effectively conveying patterns and information within the data.
A thoughtful choice of data type can enhance the user's intuitive understanding of the core information within the data.

Second, based on the data type analysis, select an appropriate chart type from the available options to visualize the resulting data for the specified task. For example, the pie chart is suitable for displaying the proportion or distribution of categorical data. Specify `label' for the categories and `value' for their corresponding values.
The line chart is appropriate for visualizing the relationship between two continuous variables across a constant range.
The user must specify the `x' and `y' columns and optionally include a `pivot\_column'.
The bar chart is used to compare discrete categories or groups.
Specify `x' for the categories and `y' for the values.
Optionally, a `pivot\_column' may display multiple series for each category.
\tichart encodes these rules within the CoT prompt, guiding the LLM to select the appropriate type of visualization chart.

Finally, LLMs verify whether the chosen chart type is appropriate among the provided options.
If inappropriate, replace it with a suitable chart type from the available options.
If no suitable visual chart is available, \tichart presents the information in tabular form.
\tichart integrates the above steps to achieve automated chart generation through a CoT-prompted LLM.

%% file: src/experiments.tex
\vspace{-0.3cm}
\section{Experimental Evaluation}  \label{sec:experiment}

In this section, we empirically evaluate the design of \name using representative benchmarks.
In particular, we focus mainly on the following research questions (RQs).

\begin{itemize}

    \item \textbf{RQ1: }How accurate is \tisql in performing text-to-SQL tasks? We focus on performance, specifically referring to execution accuracy (EX). In the SQL-based EDA system, the accuracy of SQL significantly influences the overall performance of the EDA system.
    \item \textbf{RQ2: }How effective is \name in supporting user tasks? This performance includes \textbf{relevance} - Do \name accurately identify and address the core aspects of the given analysis objective?, \textbf{completeness} - Does \name provide necessary information to address the given analysis objective?, and \textbf{understandability} - Does \name response easily interpretable, clear, and understandable for users, even with complex analysis objectives?.
    \item \textbf{RQ3: }How does latency between different components affect the effectiveness of the design in achieving the analysis objectives?
    \item \textbf{RQ4: }How do different LLMs compare in terms of costs? We focus on balancing accuracy and cost in the production environment at PingCAP.
\end{itemize}

In the following, we answer \textbf{RQ1} in Section~\textcolor{blue}{\S \ref{rq1}}, \textbf{RQ2} in Section~\textcolor{blue}{\S \ref{rq2}}, \textbf{RQ3} in Section~\textcolor{blue}{\S \ref{rq3}} and \textbf{RQ4} in Section~\textcolor{blue}{\S \ref{rq4}}.

\vspace{-0.3cm}
\subsection{Experimental Setup}

\subsubsection{LLMs}
We support prompt-based LLMs, including GPT-4-0613 (referred to as GPT-4)~\cite{gpt4}, GPT-4o~\cite{gpt4o}, GPT-4o mini~\cite{openai2024mini}, Claude-3 Opus~\cite{anthropic2024claude}, Claude-3 Sonnet~\cite{anthropic2024claude}, and Claude-3 Haiku~\cite{anthropic2024claude}. 
We utilize the \textit{text-embedding-ada-002} model of OpenAI to generate embeddings.
We set the \textit{temperature} parameter to $0$, a common configuration for the product that ensures the stability of the output. 
Furthermore, we set the \textit{top\_p} parameter to $1$.
For LLMs of OpenAI, the \textit{frequency\_penalty} and \textit{presence\_penalty} are both set to $0$.

\subsubsection{Environment}
We conducted extensive experiments with two pods on AWS Kubernetes.
Each pod is configured with 2000 millicores of CPU and 4096 MiB of memory.
The CPU model is Intel(R) Xeon(R) Platinum 8223CL, running at 3.00 GHz.
The pods run Debian GNU/Linux 12 (Bookworm).

\subsubsection{Benchmark Datasets}
In our experimental setup, we utilize three datasets, including a micro-benchmark Financial \cite{fedfunds, djia, unrate, nasdaq, gold, crude, csi300}, and two macro-benchmarks Spider \cite{spider, spider_github}, and Bird \cite{bird, bird_sql}. We maintain the Financial dataset on the TiDB cloud and automate the daily update of the database at 0:00 by retrieving the latest data.

\vspace{-0.3cm}
\subsection{Performance of \tisql (RQ1)} \label{rq1}

In this section, we evaluate the performance of \tisql using the popular text-to-SQL benchmarks, Spider and Bird.
We employ the execution accuracy (EX) metric for the Spider primarily because our main concern in practical text-to-SQL is whether the generated SQL statements yield the correct execution results.
We utilize the EX metric and the reward-based valid efficiency score (R-VES) for the Bird benchmark.
The experimental evaluation of \tisql is available at \textcolor{blue}{\url{https://github.com/tidbcloud/tiinsight}}.

\begin{table}[!t]
\caption{Execution accuracy (EX) on the test set of Spider. }
\vspace{-0.3cm}
\label{tab:spider_test}
\setlength{\tabcolsep}{0.8mm}{
\begin{tabular}{llc}
\toprule
\textbf{Classification} & \textbf{Methods} & \textbf{Test EX (\%)}\\\hline
Rule-based & Duoquest~\cite{baik2020duoquest} & 63.5 \\\hline
\multirow{4}{*}{PLM-based} & BRIDGE v2 + BERT~\cite{lin2020bridging} & 64.3 \\
 & T5-3B~\cite{scholak2021picard} & 70.1 \\
 & T5-3B-PICARD~\cite{scholak2021picard} & 75.1 \\
 & RESDSQL-3B + NatSQL~\cite{li2023graphix} & 79.9 \\ \hline
\multirow{6}{*}{LLM-based} & DIN-SQL + CodeX~\cite{pourreza2024din} & 78.2 \\
 & C3 + ChatGPT~\cite{dong2023c3} & 82.3 \\
 & DIN-SQL + GPT-4~\cite{pourreza2024din} & 85.3 \\
 & DAIL-SQL + GPT-4~\cite{gao2023text} & \textbf{86.2} \\
 \rowcolor{yellow} & DAIL-SQL + GPT-4 + SC~\cite{gao2023text} & \textbf{86.6} \\
 \rowcolor{ocean} & \textbf{TiSQL + GPT-4 (ours)} & \textbf{86.3} \\
\bottomrule
\end{tabular}
}
\end{table}

In Table~\ref{tab:spider_test}, we compare the EX metric in \tisql with those of the rule-based, PLM-based, and LLM-based text-to-SQL approaches using the Spider test dataset.
The results indicate that \tisql with GPT-4 (i.e., \tisql + GPT-4) achieves SOTA performance, reaching 86.3\%.
The DAIL-SQL employs the GPT-4 model along with the self-consistency SQL refinement, achieving an accuracy of 86.6\%.
In comparison, \tisql is only 0.3\% lower in execution accuracy than DAIL-SQL + GPT-4 + SC.
However, the self-consistency approach is time-consuming and incurs significantly higher costs than the original DAIL-SQL method.
Consequently, self-consistency is not employed in \tisql.
In the published DAIL-SQL, the original DAIL-SQL remains the primary focus; however, our \tisql achieves a performance improvement of over 0.1\% compared to the original.
It is noted that \tisql does not undergo any fine-tuning process, yet its execution accuracy still exceeds that of the SOTA PLM-based method.
This is primarily due to HDC, which aims to maximize the use of database schema information, thereby achieving full utilization of schema information in \tisql.
This further demonstrates the strong contextual learning and summarization capabilities of LLMs.
\tisql is designed to leverage the full capability of LLMs.
On the other hand, \tisql employs a combination of coarse-grained and fine-grained filtering methods for tables and columns to achieve more accurate recall.
Ultimately, the refinement chain of \tisql enhances execution accuracy.

\begin{table}[!t]
\caption{Execution accuracy (EX) on the test set of Spider using different LLMs for TiSQL.}
\vspace{-0.3cm}
\label{tab:spider_llms}
\setlength{\tabcolsep}{0.8mm}{
\begin{tabular}{lccccc}
\toprule
\multirow{2}{*}{\textbf{Methods}} & \multicolumn{5}{c}{\textbf{Test EX (\%)}} \\
 & \textbf{easy} & \textbf{medium} & \textbf{hard} & \textbf{extra} & \textbf{all} \\\hline
TiSQL + Claude-3 Opus & 91.1 & 86.8 & 80.8 & 72.5 & 84.1 \\
TiSQL + Claude-3 Haiku & 84.5 & 84.0 & 61.6 & 55.7 & 74.6 \\
TiSQL + Claude-3 Sonnet & 90.0 & 83.8 & 67.0 & 65.3 & 78.4 \\
TiSQL + GPT-4o & \textbf{91.9} & 86.2 & 79.9 & 73.1 & 83.9 \\
TiSQL + GPT-4o mini & 91.3  & 68.1 & 73.0 & 59.4 & 72.8\\
\rowcolor{ocean} \textbf{TiSQL + GPT-4} & 89.8 & \textbf{89.0} & \textbf{83.6} & \textbf{78.7} & \textbf{86.3} \\
\bottomrule
\end{tabular}
}
\end{table}

In Table~\ref{tab:spider_llms}, we give the EX of various LLMs utilized in \tisql on the Spider test dataset.
The results indicate that \tisql achieves the highest execution accuracy when utilizing GPT-4.
And, in tasks of varying difficulty, \tisql using GPT-4 has achieved SOTA execution accuracy.
It is important to note that when \tisql using the GPT-4o model, its execution accuracy outperforms all current PLM-based methods, and its execution accuracy is 2.4\% lower than that of \tisql + GPT-4.

\begin{figure}[!t]
    \centering
    \includegraphics[width=0.22\textwidth]{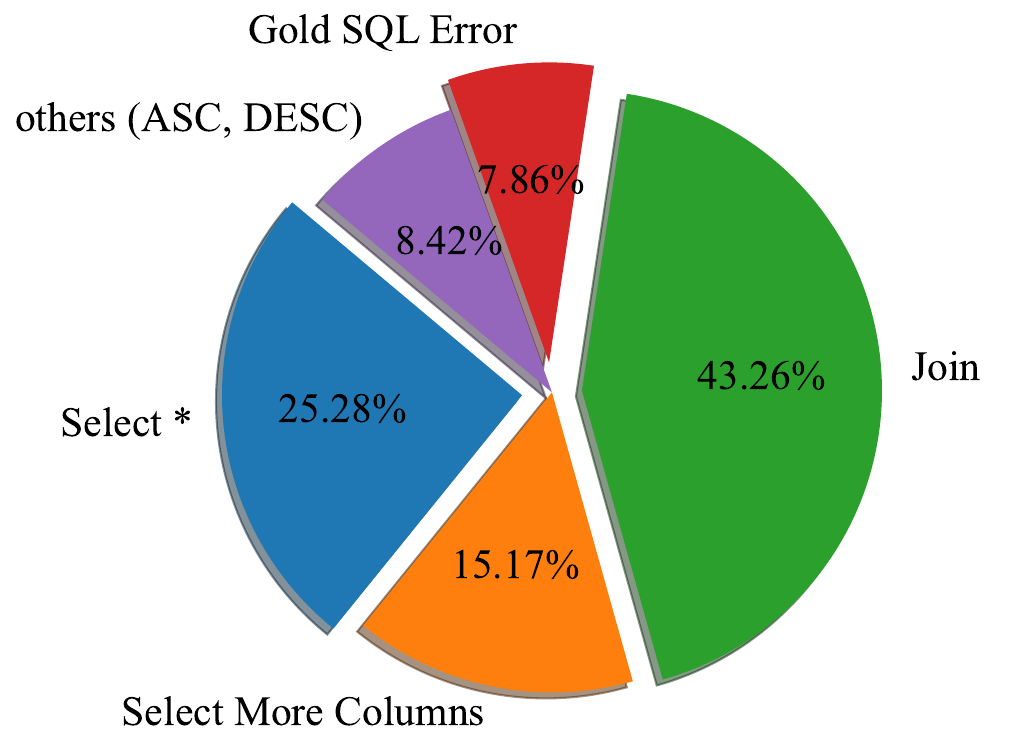}
    \vspace{-0.3cm}
    \caption{Spider dev result analysis.}
    \vspace{-0.3cm}
    \label{fig:error}
\end{figure}

We analyze all SQL statements generated on the Spider dev dataset and identify $178$ errors as false negatives, as illustrated in Figure~\ref{fig:error}.
In some cases, joining more tables accounted for the largest proportion (about 43.26\%), leading to more information being included in the final output.
There are errors in gold SQL and differences in sorting order and sorting sequence (among others).
Some \textit{SELECT} statements retrieve additional columns, or even all columns, none of which was specified in the original question.
These cases can be considered valid SQL statements in exploratory data analysis.

\begin{table}[!t]
\caption{Execution accuracy (EX) on dev/test set of Bird. Results labeled with `-' indicate that we did not submit these models to \url{https://bird-bench.github.io/} for testing; consequently, we did not report these values.}
\vspace{-0.3cm}
\label{tab:bird_ex}
\begin{tabular}{lcc}
\toprule
\textbf{Methods} & \textbf{Dev EX (\%)} & \textbf{Test EX (\%)}\\\hline
\multicolumn{3}{c}{\textbf{Baselines Methods}}\\\hline
 T5-3B~\cite{scholak2021picard} & 10.37 & 11.17 \\
 PaLM-2~\cite{bird} & 27.38 &	33.04 \\
 CodeX 175B~\cite{bird} & 34.35	& 36.47 \\
 ChatGPT~\cite{bird} & 37.22	& 39.30 \\
 ChatGPT + CoT~\cite{bird} & 36.64	& 40.08 \\
 Claude-2~\cite{bird} & 	42.70	& 49.02 \\
 GPT-4~\cite{bird} & 46.35	& 54.89 \\
 DIN-SQL + GPT-4~\cite{pourreza2024din} & 50.72	& 55.90 \\
 \rowcolor{yellow} DAIL-SQL + GPT-4~\cite{gao2023text} & 54.76	& 57.41 \\
 TA-SQL + GPT-4~\cite{qu2024before} & 56.19	& 59.14 \\
 SFT CodeS-7B~\cite{li2024codes} & 57.17	& 59.25 \\
 MAC-SQL + GPT-4~\cite{wang2024macsql} & 57.56	& 59.59 \\
 DTS-SQL + DeepSeek 7B~\cite{pourreza2024dts} & 55.8	& 60.31 \\
 \rowcolor{yellow} SFT CodeS-15B~\cite{li2024codes} & 58.47	& 60.37 \\\hline
 \multicolumn{3}{c}{\textbf{Ours}}\\\hline
 TiSQL + Claude-3 Opus & \textbf{59.2} &-\\
 TiSQL + Claude-3 Haiku & 49.2 & - \\
 TiSQL + Claude-3 Sonnet & 53.2 & - \\
 TiSQL + GPT-4o & 57.6 &-\\
 TiSQL + GPT-4o mini & 54.2 & - \\
\rowcolor{ocean} \textbf{TiSQL + GPT-4 (ours)} &  \textbf{59.1}	& \textbf{60.98} \\
\bottomrule
\end{tabular}
\end{table}

\begin{table}[!t]
\caption{R-VES on the test set of Bird. }
\vspace{-0.3cm}
\label{tab:bird_ves}
\setlength{\tabcolsep}{7mm}{
\begin{tabular}{lcc}
\toprule
\textbf{Methods} & \textbf{Test EX (\%)}\\\hline
 GPT-4~\cite{bird} & 51.75\\
 Mistral~\cite{bird} & 52.59 \\
 DIN-SQL + GPT-4~\cite{pourreza2024din} &  53.07\\
 DeepSeek~\cite{bird} & 53.25\\
 DAIL-SQL + GPT-4~\cite{gao2023text} & 54.02\\
 SFT CodeS-7B~\cite{li2024codes} & 55.69 \\
\rowcolor{yellow} SFT CodeS-15B~\cite{li2024codes} & \textbf{56.73} \\
\rowcolor{ocean} \textbf{TiSQL + GPT-4 (ours)} &  \textbf{56.06}\\
\bottomrule
\end{tabular}
}
\end{table}

In Table~\ref{tab:bird_ex} and Table~\ref{tab:bird_ves}, we present the EX and R-VES results for the Bird dataset. 
We found that \tisql outperforms DAIL-SQL + GPT-4 in all metrics.
This demonstrates that the \tisql outperforms DAIL-SQL + GPT-4 in handling complex text-to-SQL scenarios.
The performance on R-VES is slightly inferior to that of SFT CodeS-15B.
However, it is important to note that SFT CodeS-15B requires complex fine-tuning, which restricts its ability to generate SQL across diverse data domains.
The results indicate that \tisql demonstrates SOTA performance.
This finding is consistent with the results obtained from the Spider dataset.

In the public testing scenario, since the gold SQL of users is unavailable, we can only collect data on the execution success rate.
During the six months of public testing, we evaluated the execution success rate of \tisql in responding to user inquiries, achieving \textbf{82.3\%}.
The text-to-SQL tasks span various domains, including finance, retail, and gaming.

\begin{figure*}[!t]
    \centering
    \subfloat[Financial]{\includegraphics[width=0.20\textwidth]{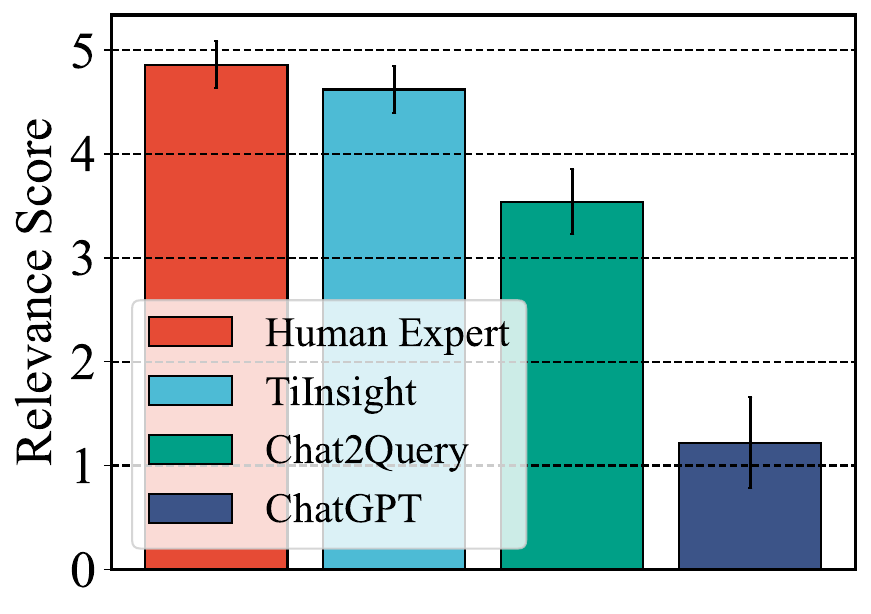}}
    \subfloat[Spider]{\includegraphics[width=0.20\textwidth]{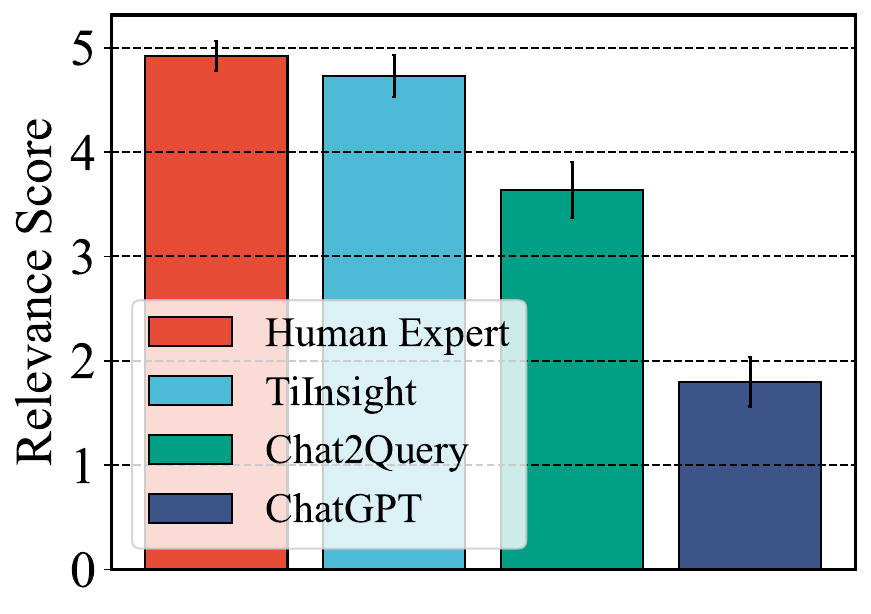}}
    \subfloat[Bird]{\includegraphics[width=0.20\textwidth]{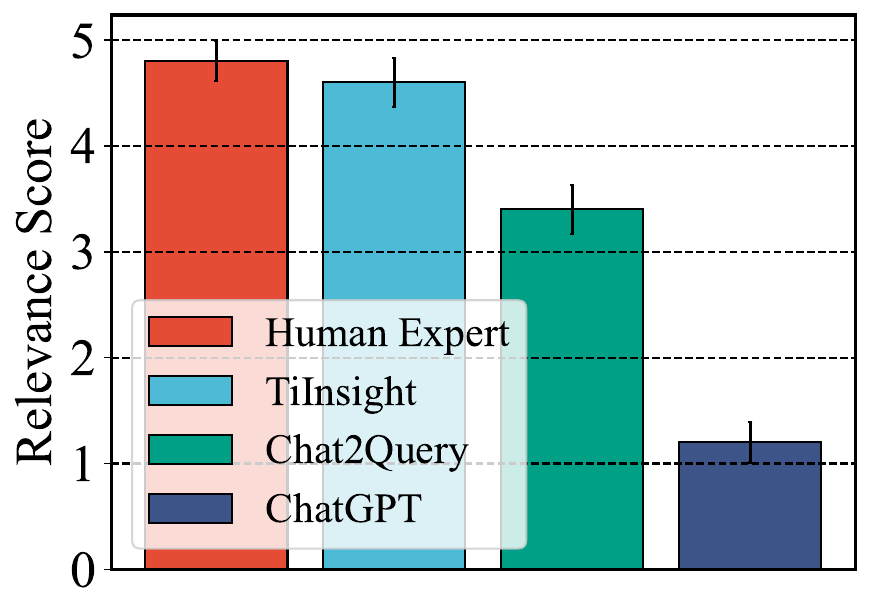}}
    \vspace{-0.3cm}
    \caption{Relevance scores of the user study.}
    \vspace{-0.3cm}
    \label{fig:1-3}
\end{figure*}

\begin{figure}[!t]
    \centering
    \subfloat[Completeness]{\includegraphics[width=0.20\textwidth]{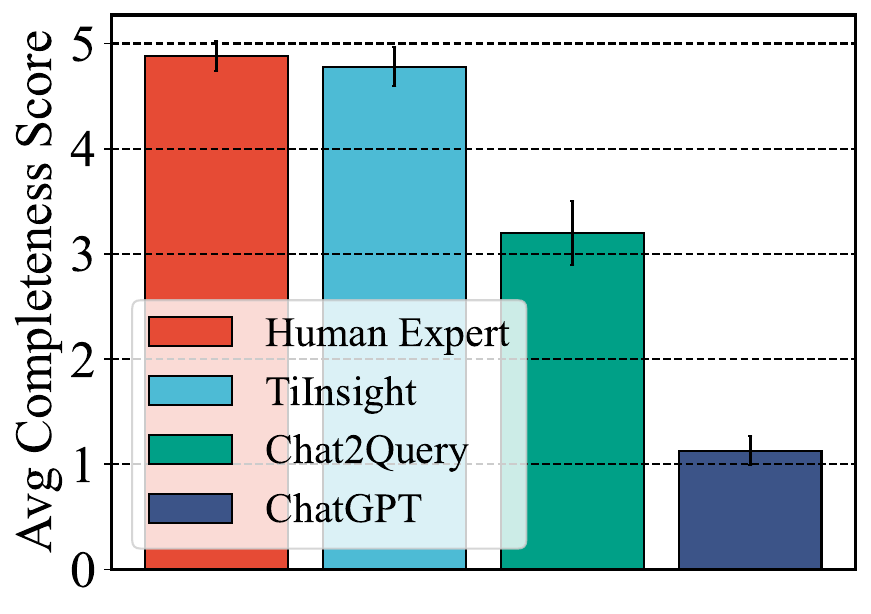}}
    \subfloat[Understandability]{\includegraphics[width=0.20\textwidth]{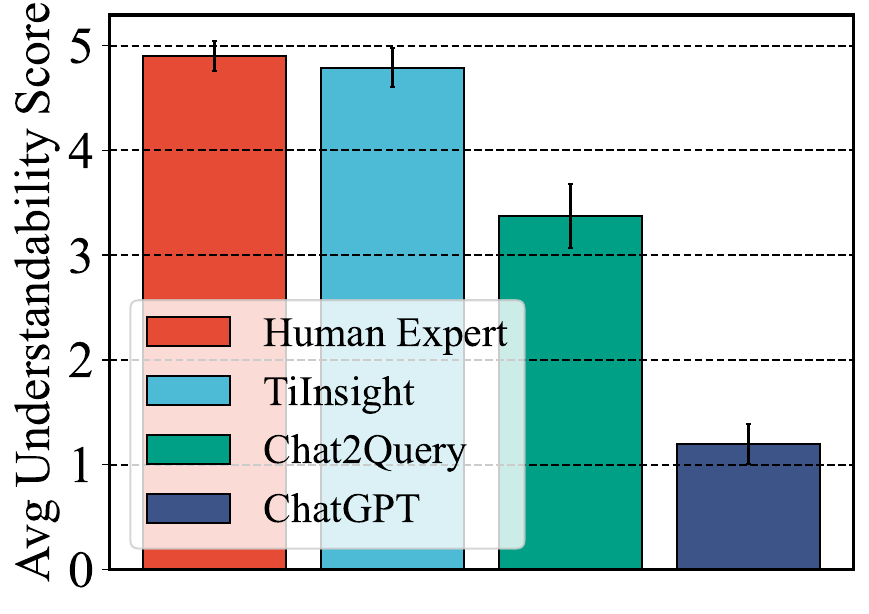}}
     \vspace{-0.3cm}
    \caption{Average completeness and understandability scores.}
    \vspace{-0.3cm}
    \label{fig:4-5}
\end{figure}

\vspace{-0.3cm}
\subsection{User Study of \name (RQ2)} \label{rq2}

\begin{figure}[!t]
    \centering
    \includegraphics[width=0.28\textwidth]{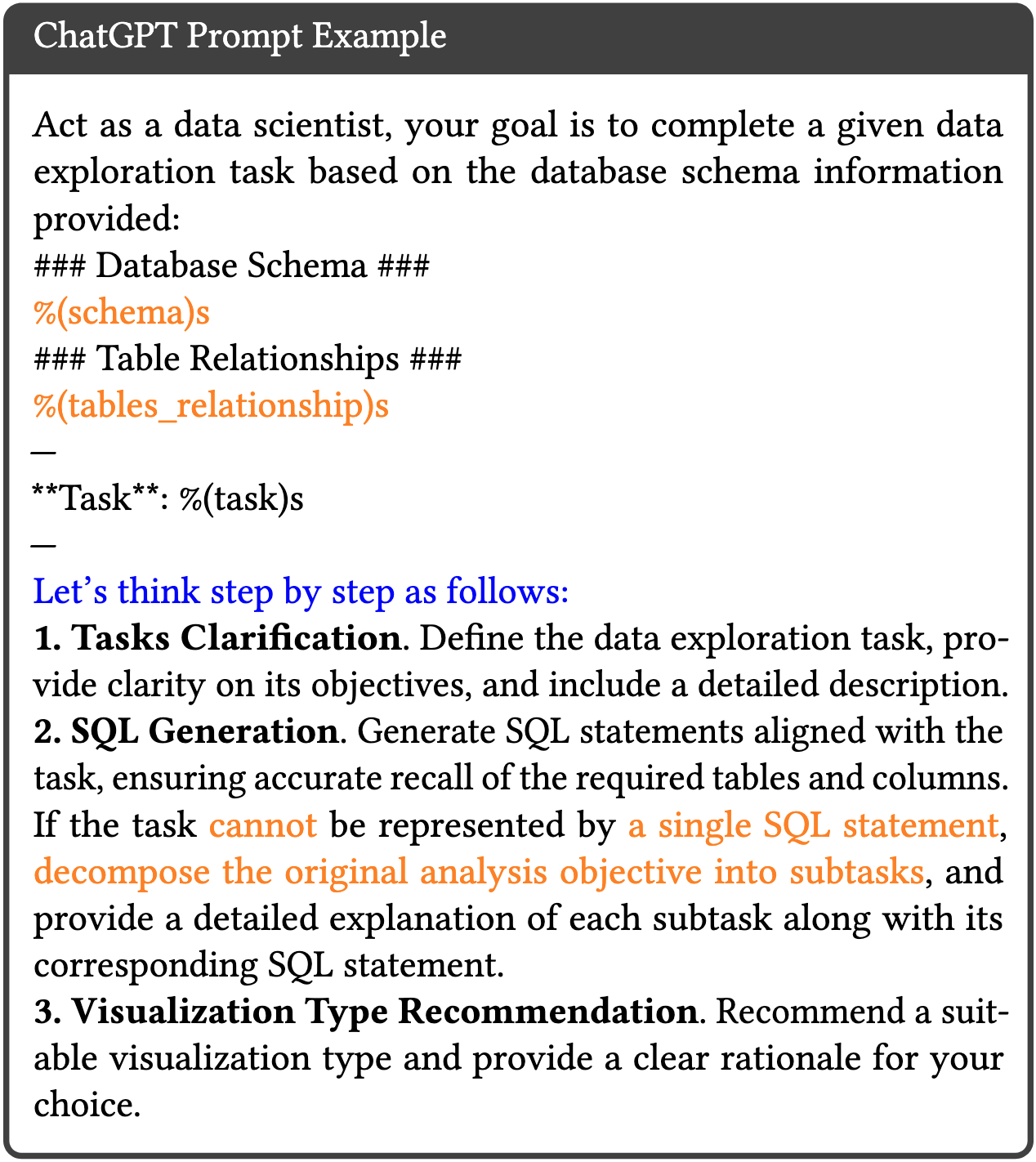}
     \vspace{-0.3cm}
    \caption{The prompt for ChatGPT.}
    \vspace{-0.3cm}
    \label{fig:prompt}
\end{figure}

We invited $20$ participants for this user study.
These individuals have data analysis experience, typically perform data analysis tasks, and are relatively familiar with SQL.
Initially, \name recommends three exploration questions to the participants, allowing them $60$ minutes to explore the three datasets, including Financial, Spider, and Bird.
Participants are asked to rate the system's relevance, completeness, and understandability on a scale from $1$ to $5$.

We evaluate \name along with the following baselines: (1) \textbf{Chat2Query} \cite{chat2query}, which we use to explore the three datasets. Chat2Query is an exploratory data analysis system that recommends appropriate data visualization representations during exploration. (2) \textbf{ChatGPT} \cite{chatgpt2022}, where we encode metadata information for the datasets in the prompt. The ChatGPT version we use cannot produce suitable visualization results, so we ask it to recommend an appropriate chart type. It utilizes a CoT prompt in Figure~\ref{fig:prompt}. (3) \textbf{Human Expert}, who uses the same exploration questions as the previous two methods. The data analysis expert manually generates the exploration results and visualizes them to establish an upper limit for the exploration results.

During the rating process, participants are specifically reminded of the following questions:

\noindent\textbf{Q1: }When tasked with analyzing datasets, do these systems accurately comprehend the user questions? In other words, must you repeatedly prompt the system to correct questions in data exploration throughout the process?

\noindent\textbf{Q2: }We require all systems to generate appropriate SQL statements during the exploration process, which may necessitate breaking down the task into sub-problems. Are the decomposed subproblems reasonable?  Do the SQL statements generated through this process meet your expectations? Does it make you feel ``wow''?  We believe that the ``wow'' factor encourages users to explore further.

\noindent\textbf{Q3: }Do you believe that the charts recommended by the system can help you understand the results of SQL queries more intuitively?

\noindent\textbf{Q4: }Do you believe that you gained a deeper understanding of this dataset during the data exploration process? This indicates that you acquired more insights through your exploration.

\noindent\textbf{Q5: }Do you find yourself attempting to ask additional questions after completing a given task? To what extent are you interested in further exploring the dataset?

The relevance scores are shown in Figure~\ref{fig:1-3} for the three datasets.
This result represents the average scores from all participants, accompanied by a 95\% confidence interval.
The human expert obtained the highest relevance score, which aligns with our expectations, as human experts are more familiar with the dataset and possess superior knowledge of data analysis.
The results also show that \name is comparable to the human expert.
This can be attributed to the design of \name, which centers around one fundamental question: How can human experts conduct data exploration?
Therefore, we propose the hierarchical data context.	
Throughout the data exploration process, \name gains more insights from datasets than other methods.

Chat2Query is slightly worse than \name, yet significantly better than ChatGPT.
Although Chat2Query is an advanced data exploration approach, it is limited by its inability to decompose complex questions into multiple sub-questions, restricting users from gaining deeper insights.
Additionally, it produces a lower SQL ``wow'' factor compared to \name.
The SQL generated by \name exhibits a more sophisticated structure compared with Chat2Query and frequently leverages table relationships to optimize complex join queries.
Participants widely agreed that the ability to adjust the data visualization results generated by Chat2Query was a key feature.

ChatGPT exhibited the worst, as users could not execute SQL statements directly during data exploration, necessitating extra effort to verify the generated SQL.
It cannot generate visualization charts and only provides text descriptions of the chart content, requiring users to utilize visualization tools to create them manually.
As SQL execution is not possible, Chat2Query is unable to recommend appropriate visualization types based on data exploration results.
Another significant issue is that it often recommends complex visualization results, which can hinder user comprehension.
During data exploration, users often need to prompt the LLM to understand the data analysis objective accurately, and the hallucination problem in ChatGPT is particularly pronounced.
This further emphasizes the importance of carefully designing the prompt for \name.
After engaging with our specified analysis goals, users often show little interest in pursuing further explorations.

Figure~\ref{fig:4-5} presents user scores for completeness and understandability, with each score reflecting the average from all participants across the three datasets.
The same conclusion can be drawn regarding completeness and understandability.
This further demonstrates that \name has been well-received by users from multiple perspectives.

The aforementioned experimental results represent subjective evaluations of the participants.
Currently, no open-source benchmark is available for the objective evaluation of exploratory data analysis systems.

\vspace{-0.3cm}
\subsection{Experiments on Latency  (RQ3)} \label{rq3}

\begin{figure}[!t]
    \centering
    \includegraphics[width=0.20\textwidth]{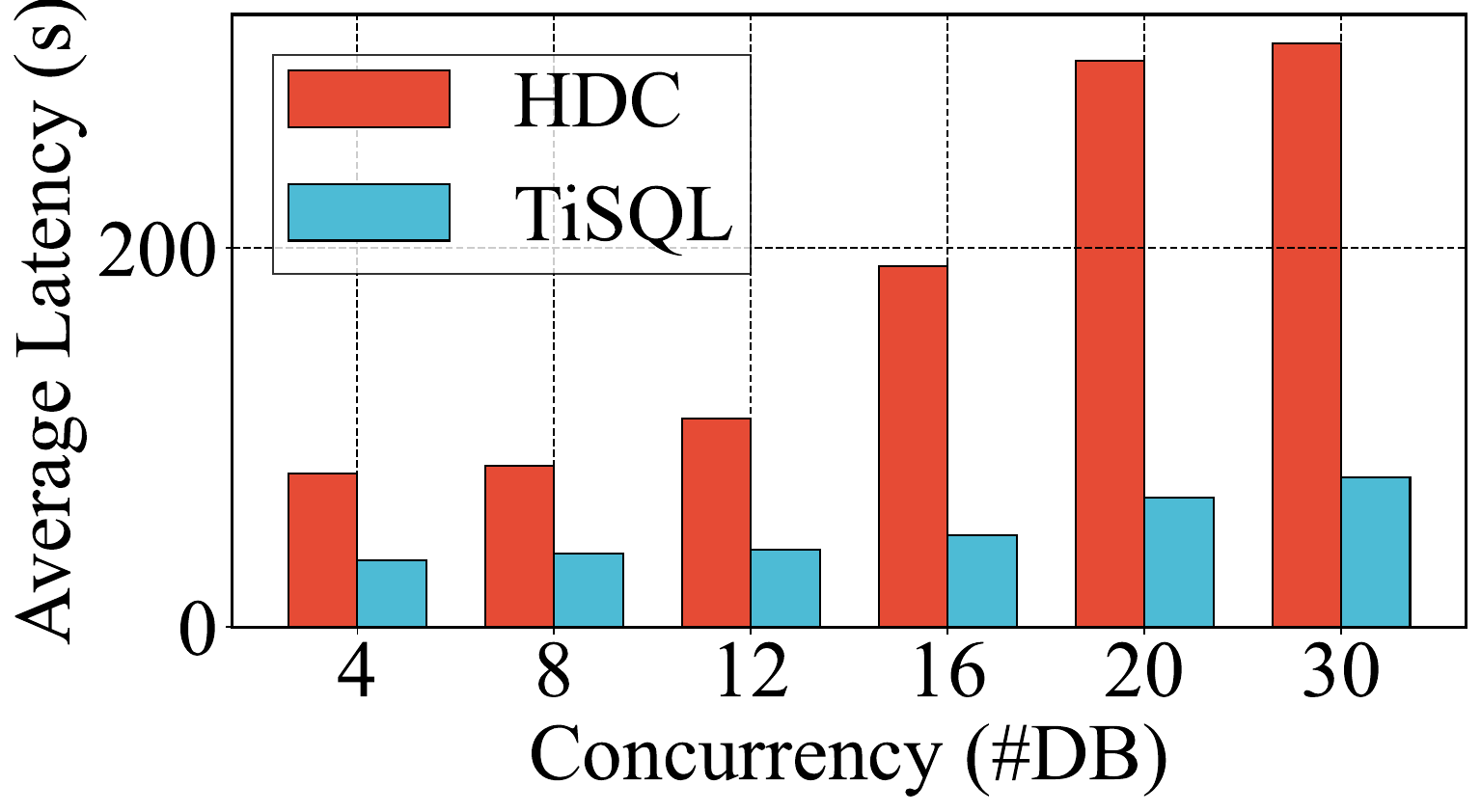}
    \vspace{-0.3cm}
    \caption{Average latency for HDC and \tisql.}
    \vspace{-0.3cm}
    \label{fig:6}
\end{figure}

We evaluate the average latency of various components of \name on the Spider dataset.
For users, average latency serves as a crucial indicator of system performance.
Figure~\ref{fig:6} depicts the average time required to generate the HDC for a database and to generate an SQL during concurrent exploration of varying numbers of databases.
The concurrency involves using N databases for HDC and SQL generation concurrently.
The average latency of HDC generation increases with the number of explored databases.
This is primarily because as the number of databases increases, more tables and columns must be explored to generate hierarchical data contexts.
The average latency of SQL generation by \tisql is also increasing, primarily because \tisql must filter a greater number of tables and columns.
It is also observed that a single HDC generation process requires more time than that of one text-to-SQL query.
However, it is important to note that HDC generation occurs only once during the entire data exploration process; thus, this time is distributed across multiple analysis targets, resulting in an acceptable average latency.
The design of \name adopts an asynchronous mode, wherein upon submission of the analysis target, \name promptly returns a bound ID to the user. At the same time, a background thread handles the binding task.
This design enhances the overall user experience.

\begin{table*}[!t]
\caption{The price characteristics of LLMs.}
\vspace{-0.3cm}
\label{tab:price}
\scalebox{0.78}{
\begin{tabular}{cccccccc}
\toprule
\multicolumn{2}{c}{\textbf{Parameters}} & \textbf{GPT-4} & \textbf{GPT-4o} & \textbf{GPT-4o mini} & \textbf{Claude-3 Haiku} & \textbf{Claude-3 Sonnet} & \textbf{Claude-3 Opus}   \\ \hline
\multicolumn{1}{c}{} & Input  & \$30.0 / 1M tokens & \$2.5 / 1M tokens \cached{\$1.25 / 1M cached} & \$0.15 / 1M tokens \cached{\$0.075 / 1M cached} & \$0.25 / 1M tokens & \$3.0 / 1M tokens & \$15.0 / 1M tokens \\\cline{2-8}
\multicolumn{1}{c}{\multirow{-2}{*}{Pricing}} & Output & \$60.0 / 1M tokens & \$10.0 / 1M tokens & \$0.60 / 1M tokens & \$1.25 / 1M tokens  & \$15.0 / 1M tokens & \$75.0 / 1M tokens\\
\bottomrule
\end{tabular}
}
\end{table*}

\vspace{-0.3cm}
\subsection{Experiments on Cost (RQ4)} \label{rq4}

\begin{figure}[!t]
    \centering
    \subfloat[Spider Test]{\includegraphics[width=0.23\textwidth]{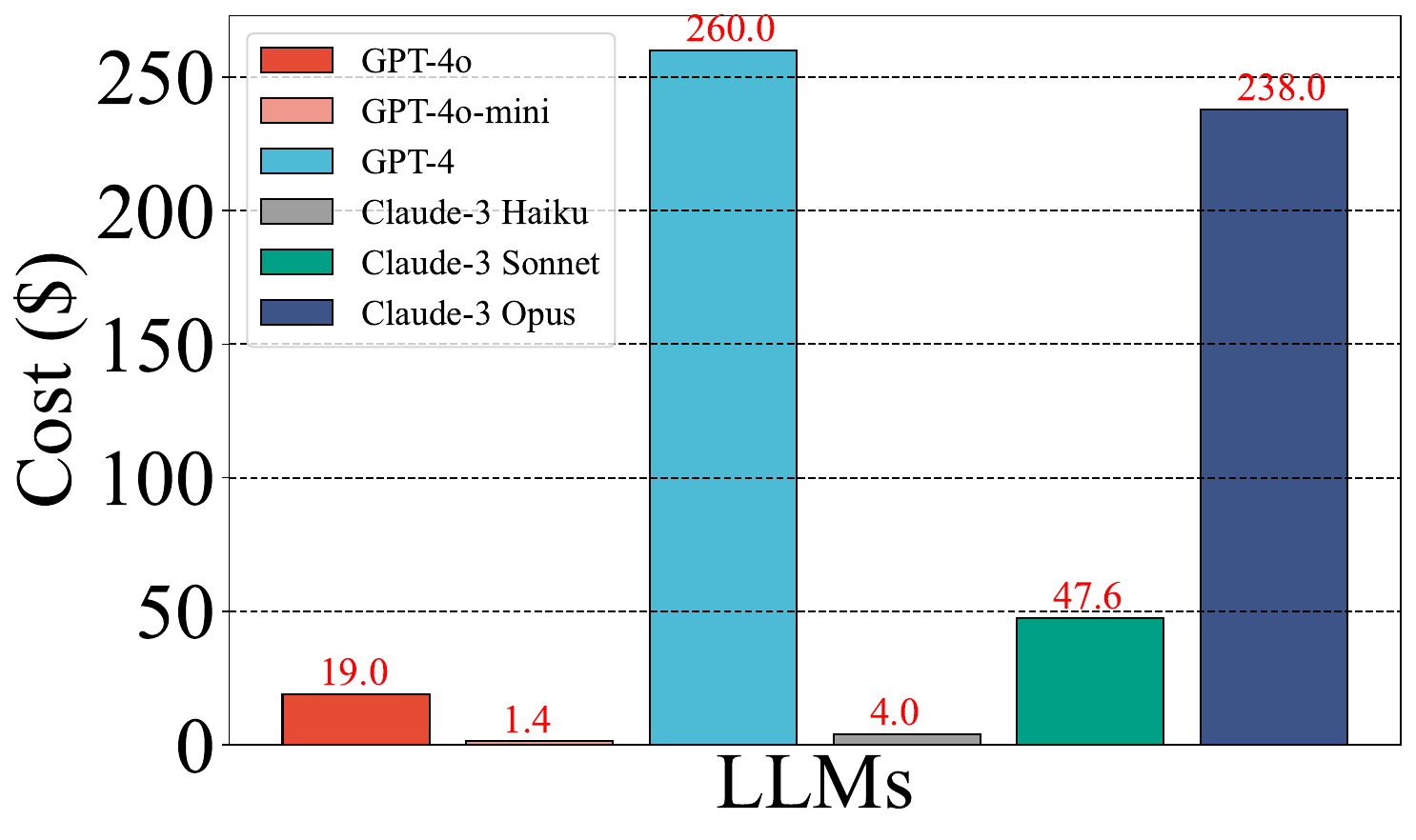}}
    \subfloat[Bird Dev]{\includegraphics[width=0.23\textwidth]{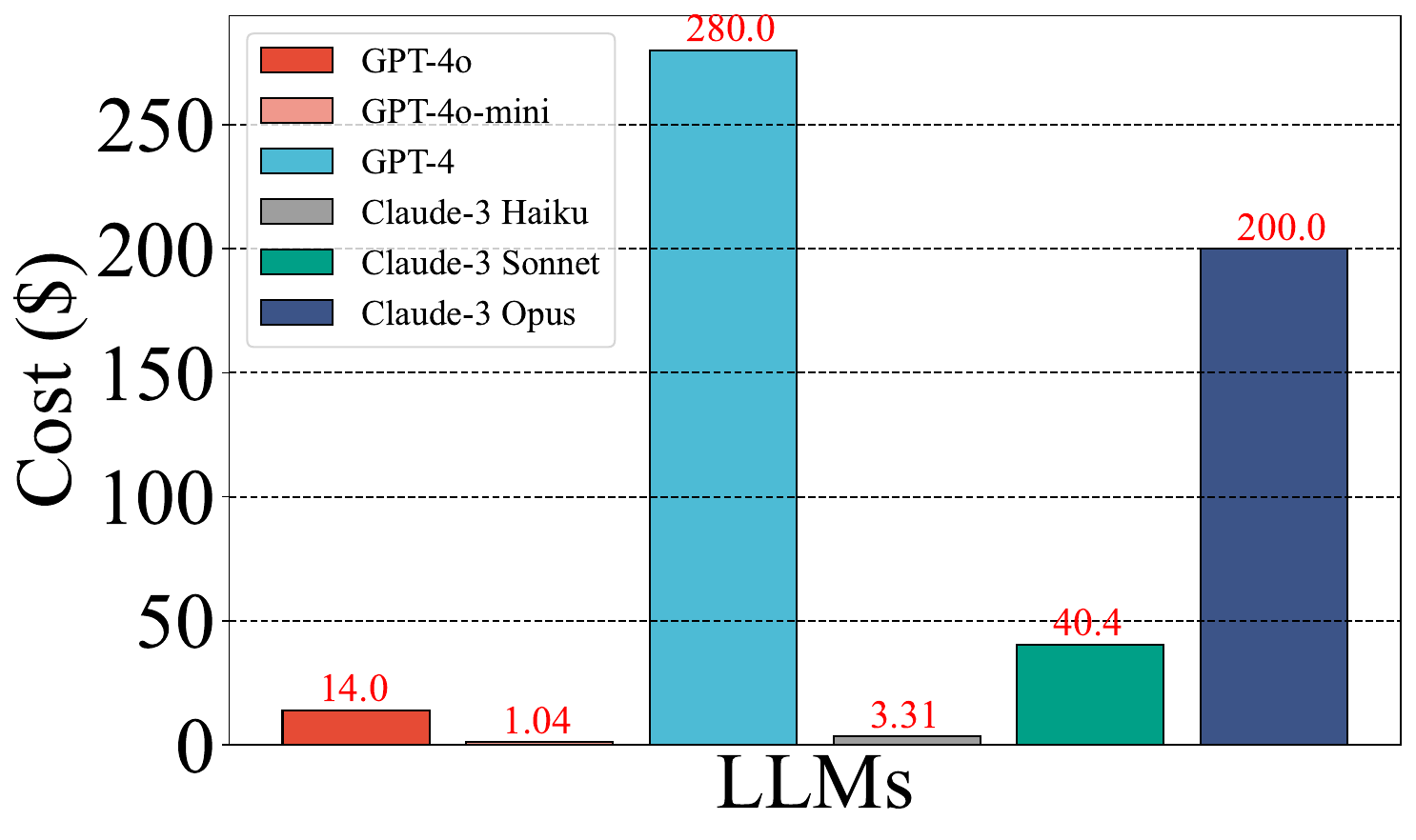}}
    \vspace{-0.3cm}
    \caption{The cost of different LLMs for \name.}
    \vspace{-0.3cm}
    \label{fig:7-8}
\end{figure}

In Figure~\ref{fig:7-8}, we analyze the average cost of \name when utilizing various LLMs across different datasets. Table~\ref{tab:price} presents the prices of the different LLMs.
The results indicate that the cost of utilizing the GPT-4 model is significantly higher (about $\$300$) than other models.
Specifically, the cost is \$0.03 per $1,000$ token for input and \$0.06 per $1,000$ tokens for output.
This is primarily due to the high costs associated with both input and output tokens for the GPT-4 model.
GPT-4o and GPT-4o mini exhibit significantly lower costs across various datasets, primarily due to the lower cost of their tokens than GPT-4.
Simultaneously, their use of a prompt caching mechanism reduces the cost by 50\% compared to requests without caching.
In particular, for GPT-4o, the cost is \$1.25 per 1M cached input tokens, and for GPT-4o mini, it is \$0.075 per 1M cached input tokens.
However, GPT-4 does not support prompt caching.
The Claude series models require a specific API setting and code adjustments to enable this prompt caching feature, which was not configured in our experiment.
This highlights the overall trade-off between cost and accuracy in \name.

%% file: src/related.tex
\vspace{-0.3cm}
\section{Related Works}  \label{sec:relate}



\noindent{\textbf{Data Exploration}}. Data exploration is a time-consuming task. 
Numerous research and commercial systems have been developed to facilitate this task~\cite{tableau, milo2020automating, powerbi, ma2023xinsight, ma2021metainsight, ma2023insightpilot, chat2query, peng2021dataprep, el2020towards, kraska2018northstar, deutch2016qplain, fariha2019example, idreos2015overview, deutch2022fedex, ding2019quickinsights, tang2017extracting, milo2018deep, amer2024intelligent}.
Some research~\cite{tableau, powerbi, kraska2018northstar} focuses on the interface within the data exploration process, implementing interactive methods to assist non-expert users.
Some research~\cite{deutch2016qplain, fariha2019example} offers user-friendly data exploration interfaces for non-expert users via query-by-example.
These systems provide user input and output examples, after which the EDA system infers the necessary query.
Advancements in text-to-SQL technology have diminished the appeal of these systems.
Several end-to-end EDA systems~\cite{milo2018deep, milo2020automating, el2020towards} utilize deep reinforcement learning to enhance data exploration.
The complexity of reinforcement learning hinders the widespread adoption of these systems in real-world industries.
DataPrep.EDA~\cite{peng2021dataprep} is a task-centric EDA system that uses Python.
FEDEX~\cite{deutch2022fedex} provides coherent explanations for the data exploration steps. 
All of these systems automatically generate insights to guide the user.
Some research~\cite{ding2019quickinsights, tang2017extracting, ma2021metainsight, ma2023xinsight} focuses on the automatic discovery of insights in multidimensional data.
These studies concentrate on exploring and mining significant patterns within the data.	
ChatGPT~\cite{chatgpt2022} can also be utilized for data exploration.	
Chat2Query~\cite{chat2query} is an SQL-based automated exploratory data analysis system through LLMs.
However, these systems cannot perform cross-domain exploratory data analysis and cannot adapt to rapidly changing business scenarios.


\noindent{\textbf{Text-to-SQL}}. Several state-of-the-art methods~\cite{katsogiannis2023survey, kim2020natural} for text-to-SQL are broadly classified into rule-based, neural network (NN)-based, pre-trained language model (PLM)-based, and large language model (LLM)-based approaches. (1) \textbf{Rule-based Approaches}. Duoquest~\cite{baik2020duoquest}, Athena~\cite{saha2016athena} and Athena++~\cite{sen2020athena++} are three representative rule-based approaches. These methods rely on pre-defined rules or semantic parsers. However, these methods are constrained in their generalization and adaptability. (2) \textbf{NN-based Approaches}. Some research~\cite{xiao2016sequence, bogin2019representing, zhong2017seq2sql, xu2017sqlnet, guo2019towards, chen2021shadowgnn, choi2021ryansql} utilize neural networks to learn the mapping from natural language to SQL queries, with the sequence-to-sequence method being the most representative.
However, the generalization capacity of these methods is constrained by the number of neural network parameters, the volume of training data, and other factors.
(3) \textbf{PLM-based Approaches}. With the emergence of pre-trained language models like BERT~\cite{devlin2018bert}, research on text-to-SQL has gradually shifted from neural networks (NN) to pre-trained language models (PLM). Numerous PLM-based text-to-SQL methods~\cite{lin2020bridging, scholak2021picard, li2023graphix, gu2023few} have garnered significant attention from both the natural language processing and database communities.
However, these methods continue to struggle with limited generalization capabilities. Addressing cross-domain text-to-SQL challenges still necessitates substantial data retraining.
(4) \textbf{LLM-based Approaches}. 
With the emergence of LLMs, LLM-based text-to-SQL methods~\cite{pourreza2024din, gao2023text, zhang2024finsql, dong2023c3, li2024codes, fan2024combining} have garnered increasing attention from both the natural language processing and database communities.
These methods investigate prompt-based and fine-tuning-based text-to-SQL techniques.
However, these approaches have yet to fully harness the capabilities of LLMs.
Additionally, these methods lack cross-domain text-to-SQL capabilities.


\noindent{\textbf{Data Visualization}}. Data visualization~\cite{qin2020making, wu2021ai4vis, shen2022towards} is a crucial component of the data analysis workflow and has received significant attention from both the data analysis and human-computer interaction communities.
SEEDB~\cite{vartak2015seedb} is a visualization recommendation engine to facilitate fast visual analysis. Voyager~2~\cite{wongsuphasawat2017voyager} is a mixed-initiative system that blends manual and automated chart specification to help analysts engage. DeepEye~\cite{luo2018deepeye} is a novel system for automatic data visualization. Table2Charts \cite{zhou2021table2charts} learns common patterns from a large corpus of (table, charts) pairs using Q-learning. Lux~\cite{lee2021lux} is an always-on framework for accelerating visual insight discovery in the dataframe workflows.
Sevi~\cite{tang2022sevi} is a data visualization system that enables Speech2Text and Text2Vis.  
HAIChart~\cite{xie2024haichart} is a reinforcement learning-based framework designed to recommend good visualizations iteratively.
None of these methods achieve end-to-end usability in exploratory data analysis systems and often exhibit excessive complexity.

%% file: src/conclusion.tex
\vspace{-0.3cm}
\section{Conclusion}  \label{sec:conclusion}

In this paper, we propose \name, a SQL-based multi-stage automated exploratory data analysis system through LLMs to facilitate real-world cross-domain scenarios.
We propose a hierarchical data context (i.e., HDC) approach to achieve cross-domain data analysis.
HDC employs a large language model to efficiently explore the database schema, including columns, tables, and database summaries, thus facilitating a better understanding of the databases.
We propose a series of algorithms combined with a map-reduce framework to enable efficient parallel HDC generation.
Building on this foundation, we propose an efficient text-to-SQL method \tisql and a rule-based data visualization method \tichart.
Extensive experiments conducted on various evaluation datasets demonstrate the effectiveness of \name.
We open source \name and its APIs to facilitate research within the data analysis community.